\documentclass[twoside,twocolumn,9pt]{article}
\usepackage{extsizes}
\usepackage[super,sort&compress,comma]{natbib} 
\usepackage[version=3]{mhchem}
\usepackage[left=1.5cm, right=1.5cm, top=1.785cm, bottom=2.0cm]{geometry}
\usepackage{balance}
\usepackage{times,mathptmx}
\usepackage{sectsty}
\usepackage{graphicx} 
\usepackage{lastpage}
\usepackage[format=plain,justification=raggedright,singlelinecheck=false,font={stretch=1.125,small,sf},labelfont=bf,labelsep=space]{caption}
\usepackage{float}
\usepackage{fancyhdr}
\usepackage{fnpos}
\usepackage[english]{babel}
\usepackage{array}
\usepackage{droidsans}
\usepackage{charter}
\usepackage[T1]{fontenc}
\usepackage[usenames,dvipsnames]{xcolor}
\usepackage{setspace}
\usepackage[compact]{titlesec}
\usepackage{url}

\usepackage{epstopdf}

\definecolor{cream}{RGB}{222,217,201}

\begin{document}

\pagestyle{fancy}
\thispagestyle{plain}
\fancypagestyle{plain}{

}

\makeFNbottom
\makeatletter
\renewcommand\LARGE{\@setfontsize\LARGE{15pt}{17}}
\renewcommand\Large{\@setfontsize\Large{12pt}{14}}
\renewcommand\large{\@setfontsize\large{10pt}{12}}
\renewcommand\footnotesize{\@setfontsize\footnotesize{7pt}{10}}
\makeatother

\renewcommand{\thefootnote}{\fnsymbol{footnote}}
\renewcommand\footnoterule{\vspace*{1pt}%
\color{cream}\hrule width 3.5in height 0.4pt \color{black}\vspace*{5pt}} 
\setcounter{secnumdepth}{5}

\makeatletter 
\renewcommand\@biblabel[1]{#1}            
\renewcommand\@makefntext[1]%
{\noindent\makebox[0pt][r]{\@thefnmark\,}#1}
\makeatother 
\renewcommand{\figurename}{\small{Fig.}~}
\sectionfont{\sffamily\Large}
\subsectionfont{\normalsize}
\subsubsectionfont{\bf}
\setstretch{1.125} 
\setlength{\skip\footins}{0.8cm}
\setlength{\footnotesep}{0.25cm}
\setlength{\jot}{10pt}
\titlespacing*{\section}{0pt}{4pt}{4pt}
\titlespacing*{\subsection}{0pt}{15pt}{1pt}

\makeatletter 
\newlength{\figrulesep} 
\setlength{\figrulesep}{0.5\textfloatsep} 

\newcommand{\topfigrule}{\vspace*{-1pt}%
\noindent{\color{cream}\rule[-\figrulesep]{\columnwidth}{1.5pt}} }

\newcommand{\botfigrule}{\vspace*{-2pt}%
\noindent{\color{cream}\rule[\figrulesep]{\columnwidth}{1.5pt}} }

\newcommand{\dblfigrule}{\vspace*{-1pt}%
\noindent{\color{cream}\rule[-\figrulesep]{\textwidth}{1.5pt}} }

\makeatother

\twocolumn[
  \begin{@twocolumnfalse}
\vspace{3cm}
\sffamily
\begin{tabular}{m{4.5cm} p{13.5cm} }

& \noindent\LARGE{\textbf{{Impulsive laser-induced alignment of OCS molecules at FERMI}}} \\
\vspace{0.3cm} & \vspace{0.3cm} \\

 & \noindent\large{Michele Di Fraia $^{1}$$^{\ast}$, Paola Finetti $^{1}$, Robert Richter $^{1}$, Kevin C. Prince$^{1,2}$, Joss Wiese$^{3}$, Michele Devetta$^{4}$, Matteo Negro$^{4}$, Caterina Vozzi$^{4}$, Anna G. Ciriolo$^{5}$, Aditya Pusala$^{5}$, Alexander Demidovich$^{1}$, Miltcho B. Danailov$^{1}$, Evangelos T. Karamatskos$^{3,7}$, Sebastian Trippel$^{3,6}$, Jochen K{\"u}pper$^{3,6,7}$, Carlo Callegari$^{1}$} \\
& \vspace{0.5cm} \\

& \noindent\normalsize{We demonstrate the experimental realization of impulsive alignment of carbonyl sulfide (OCS) molecules at the Low Density Matter beamline (LDM) at the free-electron laser FERMI. OCS molecules in a molecular beam were impulsively aligned using 200 fs pulses from a near-infrared laser. The alignment was probed through time-delayed ionization above the sulphur $2p$ edge, resulting in multiple ionization via Auger decay and subsequent Coulomb explosion of the molecules. The ionic fragments were collected using a time-of-flight mass spectrometer and the analysis of ion-ion covariance maps confirmed the correlation between fragments after Coulomb explosion. Analysis of the CO$^+$ and S$^+$ channels allowed us to extract the rotational dynamics, which is in agreement with our theoretical description as well as with previous experiments. This result opens the way for a new class of experiments at LDM within the field of coherent control of molecules with the possibilities that a precisely synchronized optical-pump XUV-probe laser setup like FERMI can offer.} \\

\end{tabular}

 \end{@twocolumnfalse} \vspace{0.6cm}

  ]

\renewcommand*\rmdefault{bch}\normalfont\upshape
\rmfamily
\section*{}
\vspace{-1cm}


+\footnotetext{$^{\ast}$ Corresponding author: michele.difraia@elettra.eu; Tel.: +39-040-375-8929/8598}%
+\footnotetext{\textit{$^{1}$ Elettra-Sincrotrone Trieste S.C.p.A., 34149, Basovizza, Trieste, Italy}}%
+\footnotetext{\textit{$^{2}$ Molecular Model Discovery Laboratory, Department of Chemistry and Biotechnology, Swinburne University of Technology, Melbourne, 3122, Australia}}%
+\footnotetext{\textit{$^{3}$ Center for Free-Electron Laser Science, Deutsches
+      Elektronen-Synchrotron DESY, Notkestrasse 85, 22607, Hamburg, German}}%
+\footnotetext{\textit{$^{4}$ Istituto di Fotonica e Nanotecnologie-CNR, 20133, Milan, Italy}}%
+\footnotetext{\textit{$^{5}$ Politecnico di Milano, Dipartimento di Fisica, 20133, Milan, Italy}}%
+\footnotetext{\textit{$^{6}$ The Hamburg Center for Ultrafast Imaging, University of Hamburg,
Luruper Chaussee 149, 22761, Hamburg, Germany}}%
\footnotetext{\textit{$^{7}$ Department of Physics, University of Hamburg, Luruper Chaussee 149,
22761, Hamburg, German}}%


\section{Introduction}
The locking of molecular axes to a laboratory-fixed frame has become a crucial tool in imaging structural dynamics in molecules \cite{Bis,Holm,Raki}. A number of techniques such as photoelectron imaging \cite{Bis,Holm}, electron diffraction \cite{Hans, Yang} and X-ray diffraction \cite{Kuepp}, have greatly benefited from the possibility of aligning molecules in space. Recently research has focused on alignment and orientation of simple molecules, and various techniques have been implemented using DC fields \cite{Bern,Loe,Reuss} or intense laser fields \cite{Corn,Sak,Vrakk,Nev,Holm2,Stap}. In the latter case, the use of a non-resonant laser field causes the formation of pendular states due to the interaction with the molecular polarizability \cite{Stap}. The laser alignment of OCS in different scenarios was recently investigated in detail with table-top laser sources demonstrating that coherent control of molecular motion can be attained \cite{Tripp}.\\
Lasers and high harmonic generation (HHG) light sources are capable of sub-femtosecond temporal pulse durations \cite{Paul,Anto, Sans}, however, the photon flux for such table-top sources is still fairly low (typically $10^6-10^{10}$ photons/pulse) making, for example, the investigation of multi-photon non-linear processes inaccessible. Free-electron laser (FEL) facilities can fill this gap (typical fluxes > $10^{13}$ photons/pulse). Moreover the free electron laser FERMI in Trieste can overcome one of the big limitations of soft X-ray table-top light sources: the tunability, required for instance to access specific resonant states in atoms and molecules. \\
Alignment at an FEL has been demonstrated previously using several approaches \cite{John, Glo, Kuepp, Kier} but for very fast rotational dynamics the jitter of SASE (Self Amplified Spontaneous Emission) FELs can be problematic, although advanced data analysis has been demonstrated to recover the details \emph{in silico} \cite{Fung}. In this respect the advantage of using a jitter-free FEL as FERMI is immediately evident with the final goal of recording molecular movies with femtosecond, sub-\AA ngstrom resolution.\\
In this paper we present experimental results that demonstrate the preparation of aligned molecules at the Low Density Matter beamline (LDM) at FERMI in Trieste. Specifically, OCS molecules in a cold supersonic molecular beam were impulsively aligned using 200 fs near-infrared (NIR) laser pulses and the resulting rotational dynamics was probed via the Coulomb explosion following multiple ionization with time-delayed XUV pulses above the Sulfur $2p$ edge, and subsequent Auger decay. An analysis of
averaged ion TOF spectra suffices to identify and characterize the alignment process. Rotational revival dynamics with the expected ~82 ps revival period is clearly observed. In addition, this experiment also benefited from an FEL upgrade of the repetition rate from 10 to 50 Hz shortening the acquisition time and allowing single-shot mass spectra. Ion-ion covariance maps were quickly extracted, in which the ion momentum correlation in the Coulomb explosion following single-photon ionization of OCS at 194 eV was resolved.

\section{Experimental}
The experiment was carried out at the LDM endstation \cite{sve,vic} of the FERMI FEL in Trieste. FERMI is a seeded FEL with unique properties such as spectral purity, tunability, low timing jitter, and  near-transform-limited pulses, described in more detail elsewhere \cite{All1,All2,Dana}. FERMI comprises two sources, FEL1 and FEL2, and the shorter-wavelength range of FEL2 was used in this experiment, operated for the first time at a repetition rate of 50 Hz. In this particular experiment, the FEL X-ray probe wavelength was set above the Sulfur 2p edge at 6.40 nm (194 eV).
\\
The LDM beamline was designed for static and time-resolved pump-probe studies of physical and chemical processes occurring in atoms, molecules, and their aggregates (clusters): a pump laser is available with few-fs time jitter relative to the FEL XUV pulse \cite{Dana} and with adjustable pulse duration and focal waist dimension.
\\
Here, the pump laser was used as the alignment laser at a wavelength of $\sim$800 nm, a pulse duration of 200 fs, a spot size with Gaussian distribution of 65 $\mu$m (standard deviation $\sigma$), and an average energy of 55 $\mu$J corresponding to a peak power density of 2 $\times$ 10$^{12}$ W/cm$^2$. At this value we observed no multi-photon ionisation of OCS. A FEL pulse energy of about 10 $\mu$J at 194 eV was used with a focal spot size of 21 $\mu$m (standard deviation $\sigma$ of a Gaussian distribution). \\
The beamline is equipped with a time-of-flight spectrometer \cite{vic} in Wiley-McLaren configuration to collect and discriminate the ionic fragments created upon ionization. A velocity map imaging spectrometer (VMI) \cite{Epp} for ions and electrons is also available. 
\\
OCS was premixed with helium with a concentration of 680 ppm and expanded into the vacuum chamber through a pulsed Even-Lavie valve~\cite{Hill} at a pressure of 36 bar. \\
The molecular beam conditions and the mechanical alignment were optimized using the ion signal of a quadrupole mass spectrometer (QMS) mounted behind the interaction region. The QMS was also used to rule out the presence of OCS dimers that could result in the creation of spurious ionic fragments.\\
\\
The coarse spatial overlap between the pump and probe pulses was found by means of a retractable fluorescent Ce:YAG screen placed in the interaction region. A coarse temporal overlap within 10-20 ps was found by means of a retractable antenna also placed in the interaction region. Spatial and temporal overlap were further refined by monitoring the ion signal from resonant two-color ionization of helium. Specifically, the FEL wavelength was set at the He $1s5p$ resonance (51.56 nm), and the NIR pulse was added, a step function was observed in the He$^+$ signal due to two-color multi-photon ionization: \\
He $ + h\nu_{FEL} \rightarrow$ He$_{1s5p} + mh\nu_{NIR} \rightarrow$ He$^+ + e^-$. This singly ionized signal was measured, and the delay between the two pulses was scanned. When the IR pulse preceded the XUV pulse no ionization occurred, otherwise the excited helium was ionized. This defined very quickly the temporal overlap with a precision of a few hundred femtoseconds. 

A further refinement of the temporal overlap, unaffected by intensity saturation effects, was obtained by setting the FEL photon energy above the ionization potential of He, and monitoring with the VMI spectrometer the intensity of the NIR-induced sidebands in the He photoelectron spectrum. This method \cite{Bou} allows the determination of the zero time delay between the pump and the probe within a few tens of femtoseconds. \\ 
In the measurements presented here, the NIR alignment laser was linearly and vertically polarized and thus parallel to the time of flight tube. Therefore, the difference in transit times-of-flight through the mass spectrometer between ions emitted in opposite directions was maximized. The action of the NIR alignment laser on OCS was monitored by detecting different ionic fragment channels after ionization above the Sulfur 2p edge with an FEL energy of 194 eV and subsequent Coulomb explosion. The Coulomb explosion occurs because the 2p hole undergoes rapid Auger decay to a doubly ionized state, with a significant branching ratio into two charged fragments\cite{Ank}. By varying the time delay of the pump and probe pulses, the rotational dynamics of OCS was observed in the temporal structure of the time-of-flight mass spectrum.\\

\section{Results and Discussion}

\begin{figure}[h]
\centering
  \includegraphics[height=7cm]{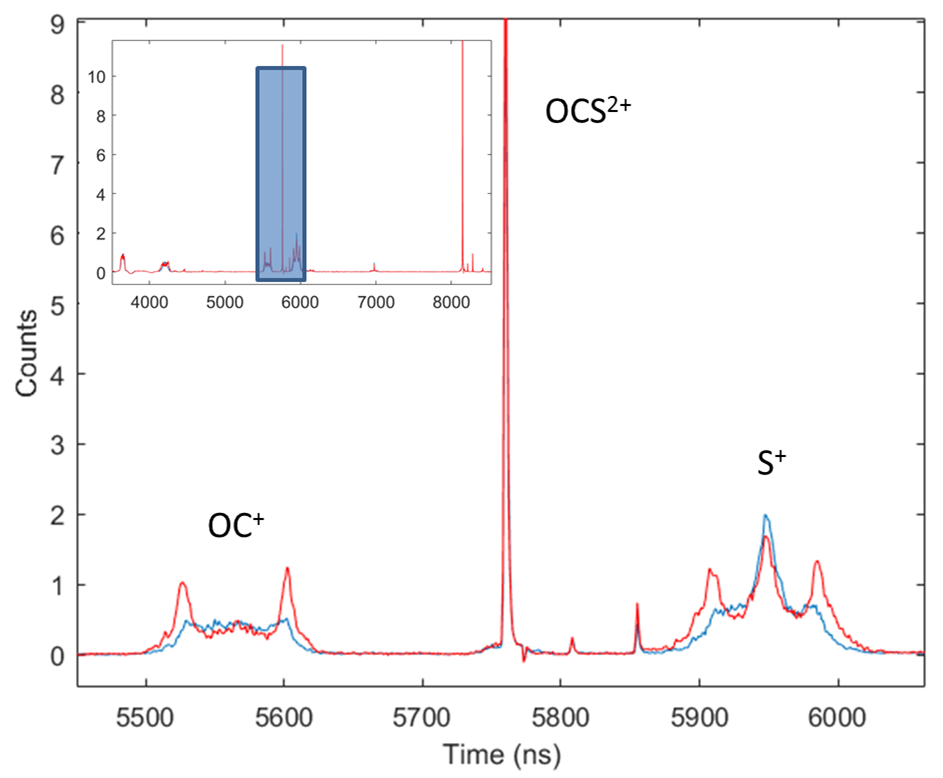}
  \caption{Time of flight spectrum of OCS with FEL pulse only at 194 eV (blue line) zoomed in the region of OC$^+$ and S$^+$ ionic fragments. Action of the NIR alignment laser (red line) with FEL pulse 39 ps after the alignment pulse. The inset is the time of flight spectrum over a larger time region.}
\label{fig:tof}
\end{figure}

In the TOF spectrum shown in Fig. \ref{fig:tof} it is possible to observe the fragmentation channels and the action of the alignment laser.

At 194 eV FEL photon energy, the TOF spectrum revealed the presence of the following ionic fragments (see inset of  Fig. \ref{fig:tof}): OCS$^+$ and CS$^+$ (longer flight times), OC$^+$, OCS$^{2+}$, S$^+$ (medium flight times), O$^+$ and C$^+$ (shorter flight times).  The spectrum observed is consistent with previous experiments performed above the Sulfur 2p edge \cite{Ank}. \\

The main part of the figure shows the fragmentation channel:
OCS + h$\nu \rightarrow$ OCS$^{2+}$ $\rightarrow$ OC$^+$ + S$^+$. 

\begin{figure}[h]
 \centering
 \includegraphics[height=7cm]{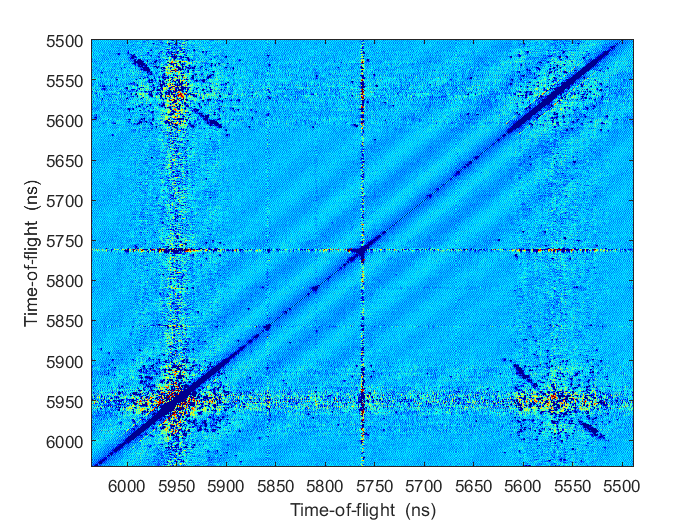}
 \caption{Ion-ion partial covariance map recorded on OCS with an FEL photon energy of 194 eV over 3000 shots. Region of OC$^+$ and S$^+$.}
 \label{fig:cov}
\end{figure}

Here the TOF distribution for each fragment is within about 90 ns and, according to SIMION$^{\textregistered}$  simulations performed on the spectrometer geometry, corresponds to a maximum kinetic energy of 3-4 eV. Thus a total kinetic energy release of the OC$^+$ + S$^+$ channel of about 7 eV is found, consistent with references \cite{Ank,Saha}.
With the FEL pulses arriving 39 ps after the NIR pulses the appearance of the two side peaks at the extremal times for each ionic fragment is a clear indication of the action of the alignment laser (red curve in Fig. \ref{fig:tof}). In fact, the NIR laser causes an alignment at half revival of the molecules along the TOF tube and the subsequent core ionization and Coulomb explosion separate the ionic fragments in two opposite directions, one towards and one away from the detector, so that the time of flight is reduced for the former and increased for the latter. In other words, the time of flight difference of the molecules exploding back to back is maximized.\\
The occurrence of Coulomb explosion following ionization is confirmed by a partial covariance map analysis \cite{Fra,Fra2} of the TOF traces reported in Fig. \ref{fig:cov}. The following partial covariance formula was used:
\begin{equation} \label{eq:pcovequation}
pcov(\textbf{X},\textbf{Y};I)=cov(\textbf{X},\textbf{Y})-cov(\textbf{Y},I) \times cov(I,\textbf{X})/cov(I,I)
\end{equation}
where $\textbf{X}$ is a time-of-flight row vector, $\textbf{Y}=\textbf{X}^T$ is the corresponding time-of-flight column vector, the first term $cov(\textbf{X},\textbf{Y})$ is the simple covariance map: $cov(\textbf{X},\textbf{Y})= <\textbf{XY}>-<\textbf{X}><\textbf{Y}>$ and the second term is the correlation term $cor(\textbf{Y},\textbf{X})=cov(\textbf{Y},I) \times cov(I,\textbf{X})/cov(I,I)$ that takes into account correlations induced by fluctuations of the FEL intensity \textit{I} and the averages are taken over $N$ FEL shots \cite{Fra}.
The covariance map confirms the correlation of the ionic fragments  OC$^+$ and S$^+$ after Coulomb explosion from the same molecule, demonstrating that this channel is a good observable for quantifying the degree of alignment of the OCS molecule. This covariance map was acquired in a time of 1 minute, for a total of $N$=3000 shots.  

\begin{figure}[!h]
 \centering
 \includegraphics[height=6.6cm]{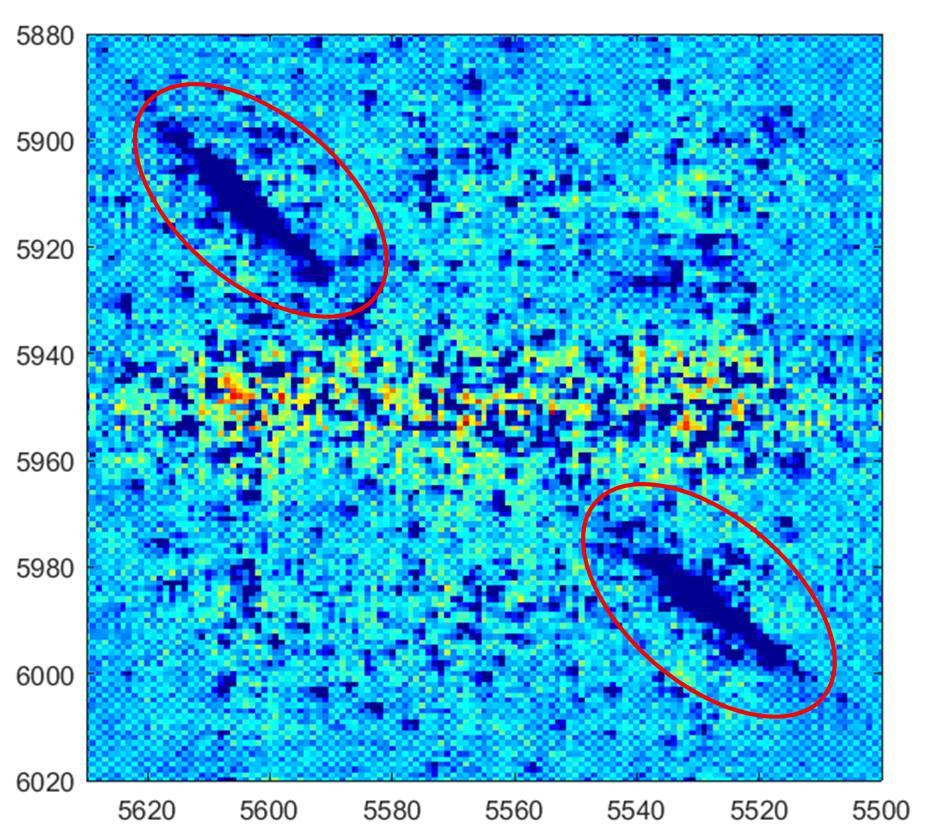}
 \caption {Partial covariance map in the OC$^+$ and S$^+$ ionic fragments region with the FEL pulse arriving after the NIR pulse with a  delay 39 ps corresponding to half revival aligned molecules. The red ellipses are guiding lines to highlight the regions where OC$^+$ and S$^+$ fragments with correlated momenta are observed.}
 \label{fig:pcov1}
 \includegraphics[height=6.6cm]{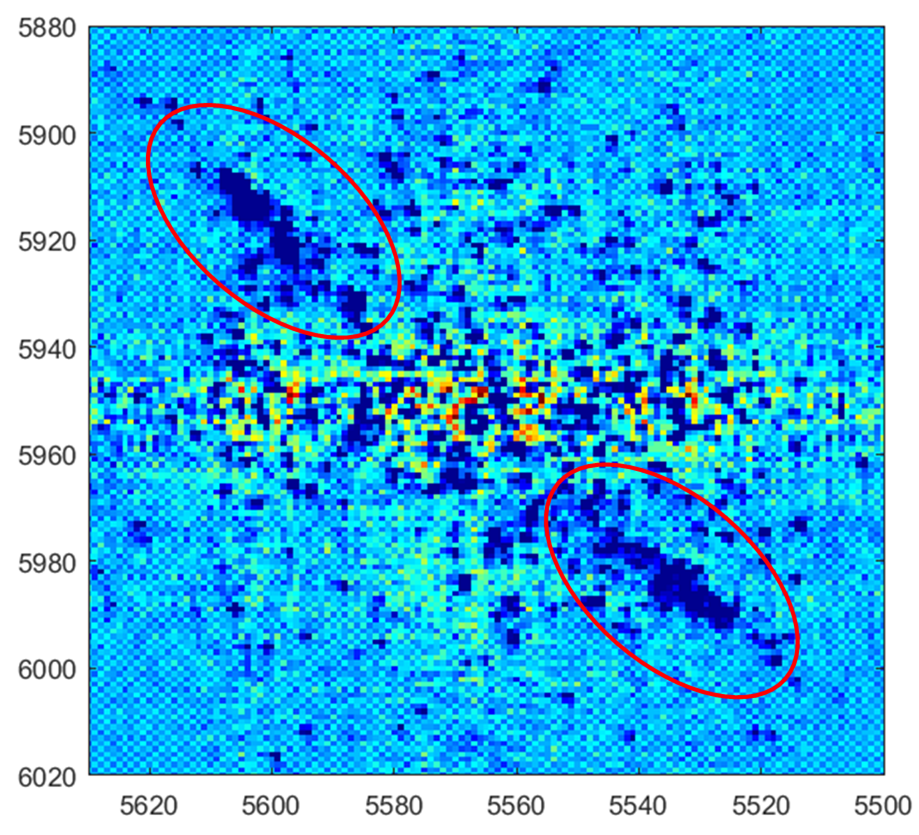} 
 \caption{Partial covariance map in the OC$^+$ and S$^+$ ionic fragments region with the FEL pulse arriving after the NIR pulse with a delay 82 ps corresponding to full revival anti-aligned molecules. The red ellipses are guiding lines to highlight the regions where OC$^+$ and S$^+$ fragments with correlated momenta are observed.}

\label{fig:pcov2}
\end{figure}

Figs. \ref{fig:pcov1} and \ref{fig:pcov2} show the region of the OC$^+$ and S$^+$ channel in the partial covariance map at two different delays between the pump and probe pulses, i.e. different average molecular directions. 

We observe a clear difference in the ionic momentum distributions when the molecules are aligned or anti-aligned with respect to the laser polarization axis.
In particular, when the molecules are anti-aligned the time of flight distribution of back-to-back flying fragments is much narrower (i.e., TOF bin values in the abscissa axis varying from 5525 to 5605 ns), and the opposite is valid at half revival for aligned molecules and the time of flight distribution is wider (i.e., TOF bin values in the abscissa axis ranging from 5515 to 5620 ns). This is due to the different flight direction with respect to the TOF spectrometer axis in the two cases, and thus to the different distribution of the ions' arrival times.\\

\begin{figure}[!h]
 \centering
 \includegraphics[height=6cm]{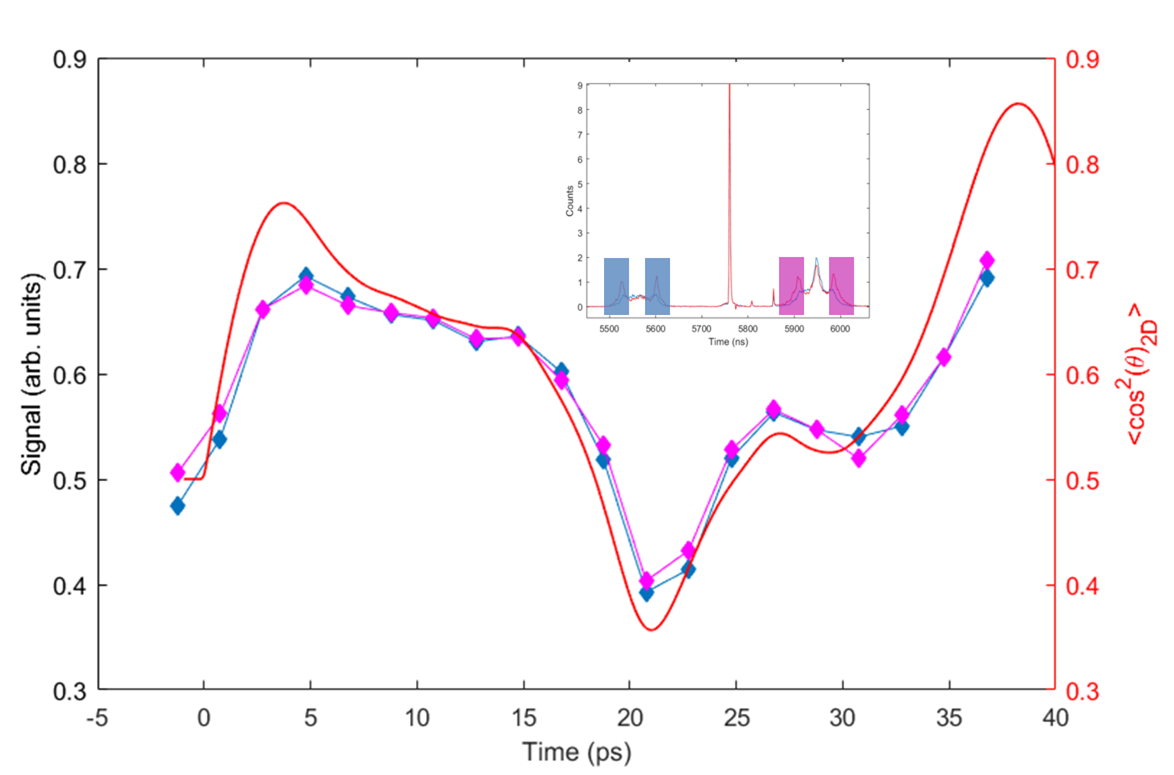}
 \caption{Rotational revival coherence dynamics monitored via the integrated signal (left axis), as highlighted in the inset with the same color code, from the OC$^+$ (blue diamonds) and S$^+$ (magenta diamonds) fragments while changing  the delay between the NIR and the FEL pulses from - 2 ps to 36 ps. The red line (right axis) is the simulated rotational revival dynamics calculated with 0.6 K rotational temperature of the molecules and 2 TW/cm$^2$ for the NIR laser intensity.}
 \label{fig:rev1}
 \includegraphics[height=6cm]{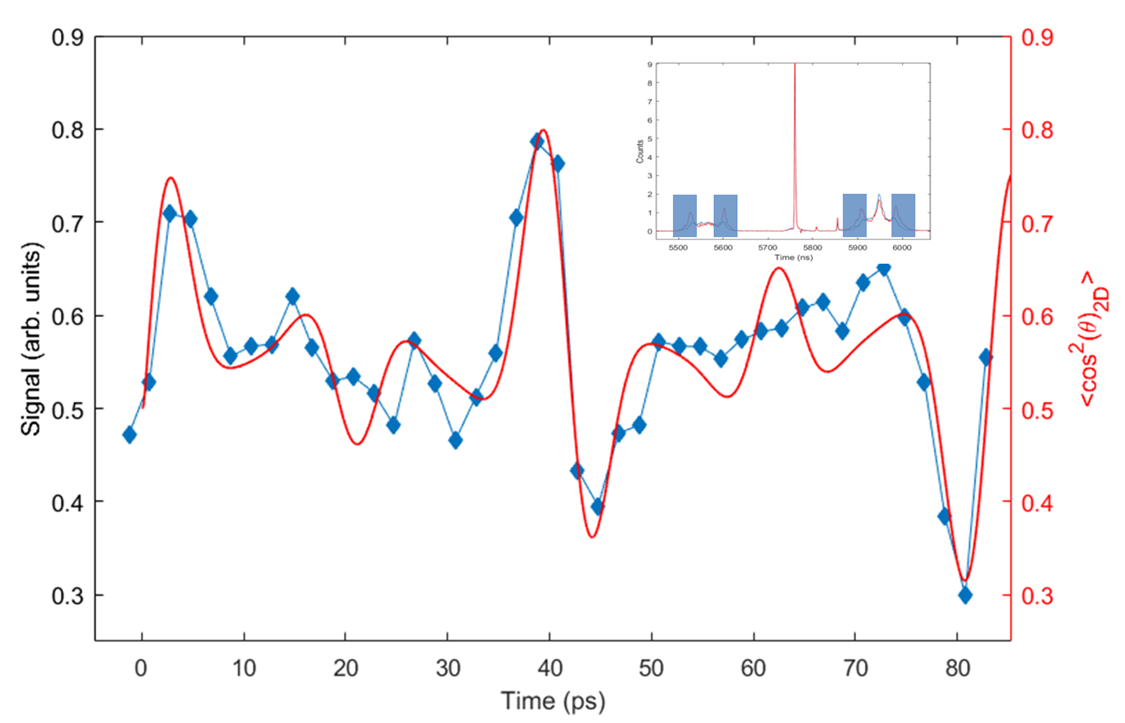}
 \caption{Rotational revival coherence dynamics monitored via the integrated signal (blue zones highlighted in the inset) as the contribution from sum of OC$^+$ and S$^+$ while changing the delay between the NIR and the FEL pulses from -2 ps to 84 ps. The red line (right axis) is the simulated rotational revival dynamics calculated with 1.5 K rotational temperature of the molecules.}
 \label{fig:rev2}
\end{figure}

Rotational revival alignment structure was also measured by monitoring the integrated signal for  OC$^+$ and S$^+$ fragments while changing the time delay of the FEL pulse with respect to the alignment NIR pulse in steps of 2 ps over a region from -2 ps to 36 ps and reported in Fig. \ref{fig:rev1}. More into details the plot shows  the integrated signal, over the shaded areas highlighted in the inset with the same color code for each fragment as in the main plot, normalized to the sum over the whole region of the inset (cutting out the region of OCS$^{2+}$).  
For each delay step, $N$=6000 shots were recorded for a total acquisition time of 36 min. \\ 
Due to a machine fault, the delay scan in Fig. \ref{fig:rev1} was interrupted and a second longer delay scan was performed with a lower gas backing pressure of 34 bar and thus a higher rotational temperature and reported in Fig. \ref{fig:rev2}.
For this second scan the degree of alignment was monitored by measuring the integrated intensities in the time-of-flight traces, as highlighted in the inset, normalized to the sum over the whole region of the inset (cutting out the region of OCS$^{2+}$). Thus the graph in Fig. \ref{fig:rev1}  shows the rotational revival dynamics for OC$^+$ and S$^+$ separately and the graph in Fig. \ref{fig:rev2} the rotational revival dynamics of the sum of OC$^+$ + S$^+$. Such rotational dynamics exhibit maxima (half revival) and minima (full revival) as a function of the delay between the NIR laser and the FEL probe. The positions in time of maxima and minima are in agreement with the positions expected for a molecule with a rotational period of 82 ps. We also found a qualitative agreement with the theoretical prediction of the quantity $\langle \cos ^2 \theta _{2D}\rangle$ (red curve) where a value of 0.8 is expected as the maximum degree of alignment for the best experimental conditions, corresponding to the lower initial rotational temperature.\\ 
The parameters used for the simulation are: NIR laser intensity of 2 TW/cm$^2$, pulse duration of 200 fs, molecular rotational temperature of 0.6 K (for the dataset in Fig. \ref{fig:rev1}) and 1.5 K (for the dataset in Fig. \ref{fig:rev2}), molecular rotational constant of 0.202 cm$^{-1}$ \cite{webCCC}, molecular population with all rotational states having rotational quantum number up to J=5 and projection M =-5,...,0,...,+5 and intensities following a Boltzmann distribution \cite{Stap}.\\
The simulations can be adjusted according to the error estimation of the main parameters: the laser energy is measured with an error < 1$\%$, the focal spot size is affected by an uncertainty in the exact focus position that is conservatively estimated to be  20$\%$, the pulse duration is measured by means of an autocorrelator with a precision of 5$\%$. The rotational temperature used for the simulation is a free parameter and adjusted to the experimental data with an error estimation of about 20$\%$, consistent with the typical rotational temperature expected using this type of pulsed valves ~\cite{Hill}. The simulations shown in Figs. \ref{fig:rev1} and \ref{fig:rev2} are the best matches to the experimental data multiplied by a re-scaling factor so that maxima and minima correspond.\\ The uncertainties in the experimental parameters result in an overall uncertainty of the simulated parameters and thus of the maximum degree of alignment achieved during this experiment. We estimated a final uncertainty for the $<$cos$^2 \theta_{2D}>$ of about 28$\%$, however, at this stage, the precision of such a quantity is not crucial for the qualitative analysis given in the present work. We point out that this work confirms that far-off-resonant ionization, followed by Coulomb explosion is an ideal, angularly unbiased probe of alignment \cite{Kuepp,Boll,Stern}. Moreover we confirm that TOF detection is capable of extracting information about molecular fragment distribution  \cite{Saha,weber} and, therefore, the evolution of the rotational wavepacket.

\section{Conclusions}
The results presented in this paper demonstrate laser induced alignment of OCS molecules in the impulsive regime, and revival dynamics were observed of ionic fragments produced after ionization and Coulomb explosion in aligned molecules. Such results demonstrate the possibility, at the LDM beamline of the FERMI Free Electron Laser, of performing future experiments where alignment of small molecules is required to allow study of photoionization processes in the molecular frame.\\
We note that LDM is also equipped with a velocity map imaging spectrometer and with the existing equipment of the beamline photoelectron angular distribution experiments on aligned molecules can be easily planned and performed in the near future.\\ 
The seeded scheme of FERMI is characterized by negligible jitter between pump and probe pulses and this property can be used in  multi-pulse alignment experiments, where an increase of the degree level of alignment due to a second NIR pulse has been already proved \cite{Post} or in multi-color (NIR + UV + FEL) pump and probe experiments, allowing access to vibrational states in aligned molecules.\\ 
Stronger alignment, as previously demonstrated for OCS \cite{Tripp, Niels}, will be required to perform experiments with photoelectron holography techniques \cite{Kra, Roll, Naka,Boll} able to image the molecules from within, and allowing the recording of molecular movies on a femtosecond scale with picometer resolution.
X-ray scattering experiments may also be performed in the future on relatively large aligned molecules, compatibly with the highest photon energies available at FERMI in the FEL2 range \cite{Kuepp, Barty}.  

\section{Acknowledgements}
The authors wish to thank the FERMI team for the support and cooperation in running the FERMI FEL light source, where this experiment was performed. J.K. acknowledges support from DESY, from the \emph{Deutsche Forschungsgemeinschaft} (DFG) through the excellence cluster ``The Hamburg Center for Ultrafast Imaging -- Structure, Dynamics and Control of Matter at the Atomic Scale'' (CUI, EXC1074), from the priority program "Quantum Dynamics in Tailored Intense Fields" (QUTIF, SPP1840), from the European Research Council through the Consolidator Grant COMOTION (ERC-K\"upper-614507), and from the Helmholtz Association ``Initiative and Networking Fund''; C.V. acknowledges the ERC Starting Research Grant UDYNI No. 307964. 



\end{document}